# An Antenna Array Initial Condition Calibration Method for Integrated Optical Phased Array

QiHao Zhang, [1,2] LingXuan Zhang, [1,2] ZhongYu Li, [1] Wei Wu, [1,2,3] GuoXi Wang, [1,3] XiaoChen Sun, [1,3,*] Wei Zhao, [1,3] and WenFu Zhang [1,3,*]

[1]State Key Laboratory of Transient Optics and Photonics, Xi'an Institute of Optics and precision Mechanics, Chinese Academy of Science, Xi'an, 710119, China
[2]University of Chinese Academy of Sciences, Beijing 100049, China
[3]School of Future Technology, University of Chinese Academy of Sciences, Beijing 100049, China
* sunxiaochen@opt.ac.cn; wfuzhang@opt.ac.cn

**Abstract:** This paper presents a modified rotating element electric field vector (modified REV) method to calibrate the antenna array initial condition of an optical phased array (OPA) device. The new method follows a similar sequential individual antenna phase calibration process while it modifies the algorithm to avoid possible π phase error in traditional REV when large initial phase distribution and finite optical power measurement accuracy present. We show that the method produces statistically more accurate and predictable calibration result which is highly desired in practice.

1. Introduction

   A variety of optical sensor technologies have been widely demanded and successfully adopted in modern electronic products from smartphones to autonomous driving systems. Optical phased array (OPA) is one of the optical technologies that frequently comes under the spotlight lately for its potential use in autonomous vehicle and 3D sensing applications. OPA uses an array of optical antennae which can be individually addressed for phase (and sometimes amplitude) tuning in order to form a beam of light and to steer the beam to desired directions. Compared to radio frequency phased array that has been extensively used in many long distance ranging applications, optical phased array hasn't seen wide adoption partly due to the practical difficulty to pack a large number of individual optical antennae (and corresponding phase tuning elements) at distances comparable to optical wavelength. With the rise of integrated photonics, Si photonics in particular, people find ways to integrate many diffraction-limited optical elements close to each other at large scale by low-cost semiconductor wafer level process. Si photonics-based integrated OPA quickly becomes a hot topic in both academic research [1-7] and entrepreneurial endeavors [8].
   Integrated OPA is in its infancy for practical applications and many engineering challenges limiting its performance and practicality await for better solutions. One such challenge is to calibrate the almost random phase error (and to a lesser extend the amplitude error) among an array of waveguide-distributed antennae due to unavoidable processing errors (such as line edge roughness and lithography errors). Even very small width variations that modern semiconductor processes provide can cause large phase error and chaotic far-field patterns when such variations are accumulated throughout millimeter long waveguide routes which are required for even moderate number of antennae. Therefore, calibrating initial phase condition of an antenna array is required for every and each OPA product and the effectiveness and efficiency of the calibration is a key to successful commercialization.
   Some mature calibration methods used in radio frequency phased array such as near-field method [9] and mutual coupling method [10] which require measuring near field or single antenna data is not practical for OPA due to its micrometer size antenna. Phase toggling method [11] demands large and fine far field pattern measurement which makes corresponding optical calibration system complex and compute-intensive. Rotating element electric field vector (REV) method [12, 13] is a more efficient far-field method while suffers possible π phase error when initial phases span the whole 2π space (which is common for OPA while less common in radio frequency) at finite optical power measurement accuracy. Other generic optimization methods such as hill climbing, genetic algorithm, particle swam optimization, stochastic gradient descent algorithm and etc. [14-17] can provide fairly good calibration in some cases however they generally cannot avoid local extremum uncertainty (which is a severe issue for most OPA with larger-than-wavelength-spaced antennae) and are not scaled well with the size of the antenna array. Here we propose a modified rotating element electric field vector (modified REV) that avoids the possible π phase error in traditional REV. We believe the improvement in our method has great implications in practical deployment of OPA products. In this paper, we describe this modified REV method, apply it to a 9x9 OPA with unevenly spaced antenna array (for better side lobe suppression) and show that it produces statistically more accurate, predictable and converged calibration result.

## 2. Theory

2.1 Modified REV

Rotating element electric field vector method (REV) calculates the initial phases of each antenna by finding the maximum and minimum optical power at a fixed single far-field point while varying the phase of each antenna from 0 to 2π. For each antenna, from the measured phase shift Δϕ from the initial state to the maximum power state, the corresponding maximum power and the minimum power values, the phase shift from its initial state to the calibrated state (i.e. a ideal phase distribution for all antennae to maximize the power at the given far field point) of this antenna is calculated. Reset this antenna to its initial phase and repeat this process for the next antenna till the phase shift of all antennae is determined and the OPA is considered calibrated. The phase shift calculation in this method is correct in theory however in practice, as explained later, can produce π phase uncertainty error due to wide initial error range (0-2π) and finite optical power measurement accuracy.

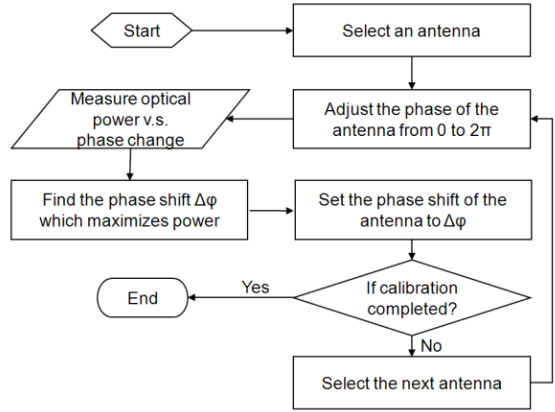

Fig. 1 Phase calibration flow of the presented modified REV method

Instead, in our modified REV method, we let Δϕ as the phase shift from the initial state to the calibrated state and set the antenna to this maximum power phase when repeating the same calibration step on the next antenna. We prove in both theory and simulation that this process quickly converges and approaches to the ideal calibration condition with the number of calibrated antennae. And this simple approach avoids π phase uncertainty and do not rely on precise maximum and minimum power measurement data. The calibration process is shown in Fig. 1 and the theory is explained in details below.

The electric field of the nth antenna at a fixed far-field point Q is

$$\mathbf{E_n} = A_n(\alpha, \beta, r_n)\exp(i\omega t - i\mathbf{k_n} \cdot \mathbf{r_n} + i\varphi_n) \cdot \hat{\mathbf{e}} \tag{1}$$

, where α, β and $r_n$ is the coordinates of point Q in polar coordinate system, $A_n(\alpha, \beta, r_n)$ is the amplitude of the antenna, $\omega$ is the optical frequency, $\mathbf{r_n}$ is the distance vector from the antenna to the point Q, $\varphi_n$ is the initial phase of the antenna, and $\hat{\mathbf{e}}$ is unit vector determined by polarization.

In practice the interested far field target is significantly farther in distance than the size of the OPA thus the directions of electric field and distance vectors of all antennae can be approximately considered the same. In addition, as all antennae share a single coherent light source the polarization of every antenna is the same. Equation (1) can then be simplified to scalar form. The total electric field at point Q is

$$E_0 = \sum_{n=1}^{N} A_n \exp(i\omega t - ikr_n + i\varphi_n) = \exp(i\omega t) \sum_{n=1}^{N} A_n \exp(i\phi_n) \tag{2}$$

, where $\phi_n = \varphi_n - kr_n$ is the phase of the nth antenna at point Q. The intensity is

$$I_0 = |E_0|^2 = |\exp(i\omega t)|^2 \cdot \left|\sum_{n=1}^{N} A_n \exp(i\phi_n)\right|^2 = \left|\sum_{n=1}^{N} A_n \exp(i\phi_n)\right|^2 \tag{3}$$

When the phase of the mth antenna is changed by $\Delta\varphi$, the intensity becomes

$$I(\Delta\varphi) = \left|\sum_{n=1}^{N} A_n \exp(i\phi_n) - A_m \exp(i\phi_m) + A_m \exp(i\phi_m + i\Delta\varphi)\right|^2 \tag{4}$$

Define

$$E_{-m} = \sum_{n=1}^{N} A_n \exp(i\phi_n) - A_m \exp(i\phi_m) = A_{-m}\exp(i\phi_{-m}) = v_m + iw_m \tag{5}$$

, where $E_{-m}$ represents the sum of the electric fields of all antennae except the mth, $A_{-m}$ is the amplitude of $E_{-m}$, and $v_m$ and $w_m$ are the real and imaginary part of the $E_{-m}$, respectively.

The intensity $I(\Delta\varphi)$ can then be expressed as

$$I(\Delta\varphi) = v_m^2 + w_m^2 + A_m^2 + 2v_m A_m \cos(\phi_m + \Delta\varphi) + 2w_m A_m \sin(\phi_m + \Delta\varphi) \tag{6}$$

A derivative calculation shows the following condition must be met at an extremum of $I(\Delta\varphi)$:

$$\Delta\varphi + \phi_m = \begin{cases} \arctan\dfrac{w_m}{v_m} \\ \pi + \arctan\dfrac{w_m}{v_m} \end{cases} \tag{7}$$

The intensity at Q reaches its maximum when $\Delta\varphi + \phi_m = \arctan(w_m/v_m)$ as shown by the second derivative of $I(\Delta\varphi)$ without losing generality by assuming $A_m > 0$:

$$\left.\frac{d^2 I}{(d\Delta\varphi)^2}\right|_{\Delta\varphi = \arctan\frac{w_m}{v_m} - \phi_m} = -2A_m\sqrt{v_m^2 + w_m^2} < 0 \tag{8}$$

And the maximum intensity $I_{max}$ is

$$I_{max} = I\left(\arctan\frac{w_m}{v_m} - \phi_m\right) = \left(\sqrt{v_m^2 + w_m^2} + A_m\right)^2 = (|E_{-m}| + |E_m|)^2 \tag{9}$$

It simply implies that the intensity of point Q reaches its maximum with the mth antenna and the combined field of the rest of the antennae are in-phase. This simple principle guides the whole calibration process. After the Δϕ of each antenna is obtained by varying its phase till maximum measured optical power at Q, the antenna is fixed at this phase and the next antenna is calibrated to the next maximum intensity condition. Repeat this process until all antennae are calibrated and set to their maximum intensity phase. The principle also guarantees the monotonic increase of the measurement optical power which helps reduce any possible ambiguity during the process. This method can be used for calibrating an OPA antenna array distributed in any patterns and does not demand any particular order in selecting antennae.

The whole process may be explained more intuitively in Figure 2 (a-f). The related electrical field is drawn in a vector plane where the length and the angle from x-axis of a field represent its absolute amplitude and phase respectively. In following text, we refer to a field vector in this representation as opposed to a real space representation.

In practice when measurement error inevitably present, the final antenna array phase condition as well as the maximum intensity at Q may still be slightly off from their ideal values however following analysis shows the calibration well approaches the ideal condition and as we later show traditional REV also present similarly if not worse results. In practice, the calibration process in production demands a statistically predictable and converged result.

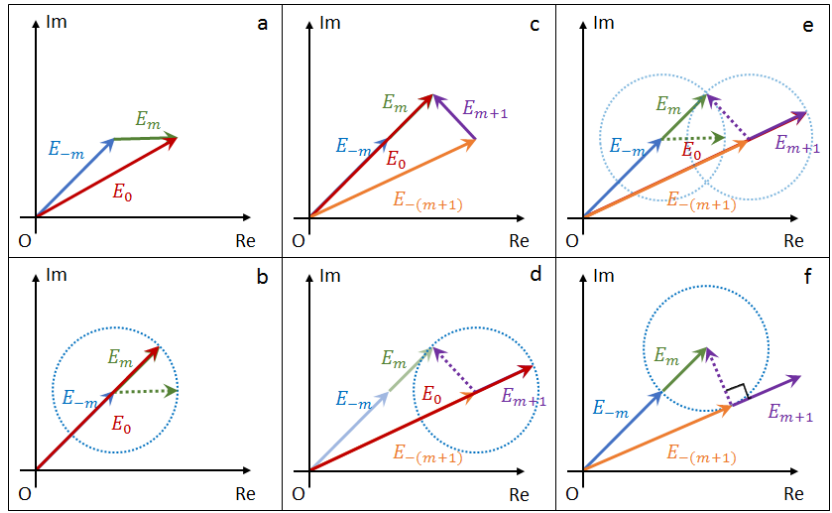

Fig. 2 (a) Electrical field decomposition at a fixed far field point Q before the mth antenna is calibrated; (b) Maximizing the total intensity at Q when $E_m$ aligns to the sum of the rest of the antennae field; (c) Keeping $E_m$ at this phase, repeat the intensity maximization process with the (m+1)th antenna; (d) The (m+1)th antenna's calibration by the same method; (e) Field decomposition history after both the mth and the (m+1)th antennae are calibrated; (f) A particular case used in error analysis where the phase change of $E_{m+1}$ after calibration is π/2.

2.2 Error Analysis

For practical considerations, we need to analyze errors resulted in this method. It can be seen that the calibrated field vectors of the mth and the (m+1)th antennae that are aligned to $E_{-m}$ and $E_{-(m+1)}$, respectively, as shown in Fig2 (e) are generally not aligned with each other. Therefore at least one of the calibrated phase values of mth and (m+1)th antennae differs from its ideal value and it represents an intrinsic error of the method. However, as we explain below such error does not pose significant overall calibration error and we also show later that it's more limited or converged than that in traditional REV method when measurement errors exist.

As shown in Fig 2 (f), the relative intrinsic phase error of a certain antenna (m+1) reaches its maximum in this case and the error, the phase difference between the (m+1)th and the mth antenna, in general satisfies

$$\delta\varphi_{m+1} \leq \arcsin\frac{|A_{m+1}|}{|E_0|} \tag{10}$$

, where $A_{m+1}$ is the amplitude of the (m+1)th antenna; $E_0$ is the combined electric field at point Q before calibrating the (m+1)th antenna. It implies $\delta\varphi_{m+1}$ continues to decrease with the increase of the number of the calibrated antennae as $|E_0|$ becomes larger when more antenna field vectors getting approximately aligned.

For example, after the first antenna is calibrated, $|E_0| \geq A_1 \cong A_2$, the phase difference between the second and the first antenna

$$\delta\varphi_2 \leq \arcsin\frac{A_2}{|E_0|} \leq \pi/2 \tag{11}$$

While for the third antenna to the calibrated, $|E_0| \geq \sqrt{2}A_3$, the phase difference

$$\delta\varphi_3 \leq \arcsin\frac{A_3}{|E_0|} \leq \pi/4 \tag{12}$$

And it can be easily seen that $\delta\varphi$ continues to reduce as more antennae are calibrated.

In an OPA where the phase error is mostly introduced by waveguide size variations, antennae initial phases tend to be widely distributed thus the intrinsic phase error after calibration also randomly distributes within $\pm\ \delta\varphi$ and the calibration imperfection caused by such distributed error is generally far less severe than the worst case scenario. In addition, if higher calibration accuracy is required, the same process can be repeated. And since $|E_0|$ becomes significantly larger after the first calibration process the intrinsic error becomes even smaller. Furthermore it is generally sufficient to only perform the second calibration on first few antennae whose phase errors are relatively larger than others.

Another type of error rises from errors in measurement (mainly measured optical power) and in calculations using the measurement data (phase change between initial to maximum power conditions). In traditional REV, such measurement error becomes more severe due to the way of calculating phase values that can introduce possible π phase error. The equation to calculate the initial phase shift of an antenna in traditional REV method is

$$\phi_e = \arctan\left[\sin(\Delta\varphi_{pmax})/\left(\cos(\Delta\varphi_{pmax}) + \frac{\sqrt{p_{max}/p_{min}} - 1}{\sqrt{p_{max}/p_{min}} + 1}\right)\right] \tag{13}$$

, where $\Delta\varphi_{pmax}$ is the phase shift from the initial state to the maximum power state; $p_{max}$ and $p_{min}$ are the maximum and minimum power at point Q, respectively.

The arctan function produces a value between [-π/2, π/2] while the possible true phase difference ranges a whole 2π. It requires the algorithm to decide whether or not to purposely add a π to its solution based on the value of $\Delta\varphi_{pmax}$. Such decision in principle has no problem, however, it presents possible error when the phase change from its initial value to the measured maximum power point occurs close to ±π/2 at which small optical power error may produce completely wrong determination.

As we show in the next section, even a 0.1% intensity error can cause poor calibration results. In a microwave radar, where the initial phase shifts of antennae are generally not as widely diverged as in an OPA, such issue is not as severe. We consider such practically inevitable error in the next section and compare its effect between traditional REV and modified REV methods.

3. **Simulation**

3.1  Description

To test our modified REV calibration method we perform the following simulation work with a 9x9 two-dimensional optical phased array operating at 1550nm wavelength. As shown in Fig 3 (a), the spacing between the antennae in both directions increases linearly from 10μm to 16μm in order to suppress side lobes. We assume all antennae share the same emission profile including intensity and far-field distribution. In another work we show that the intensity of each antenna has a much less effect than the phase on overall beam forming performance. In other words, the optical power non-uniformity introduced by some Si photonics integrated OPA designs (e.g. power splitting [18, 19] or cascaded power tapping [20, 21]) has tolerable detrimental effect. As shown in Fig. 3(b) we choose a far field point Q at 1m distance which is much farther than the distance between any two antennae so our scalar electrical field approximation holds. For convenience the Q is selected as the (0, 0, 1) point in Fig. 3(b) while any far field point is equivalent for the purpose of calculation by this method.

To simulate a real-world process of finding the maximum optical power at Q and calculating the phase shift, we recognize that the phase can only be varied in a finite step provided by the resolution of a digit-to-analog converter (DAC). We consider a DAC resolution ranges from 3-bit to 8-bit (for 0-2π range) and test its effect on calibration result.

We introduce a small Gaussian-distributed error to the far field optical power measurement data. In practice, this error rises from multiple sources including IR camera measurement inaccuracy, side lobe light scattered by a test system, scattered light from chip-coupled light source, stray ambient light, etc. We choose the Gaussian parameter of this error such that the errors are within ±0.1% (at 3σ) of the peak far field intensity to reflect its small noise nature. However as we can see later it has a non-negligible effect on calibration result.

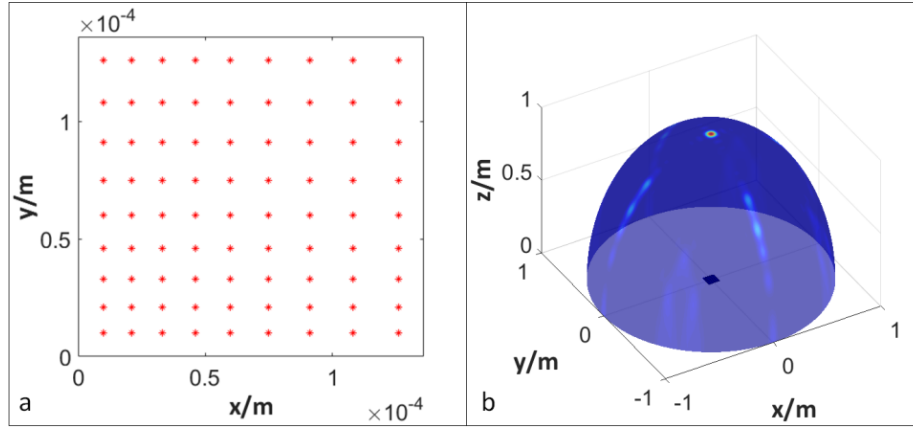

Fig. 3 (a) Schematic diagram of the antenna array. The red asterisks represent antennae. (b) The blue surface is the far field detection sphere. The color on the sphere represents the intensity when the array is ideally calibrated.

3.2 Results and Analysis

We first calculate the far field of the OPA with ideal initial phase condition, i.e. phases are equal for all antennae as shown in Fig. 4 (b) and the result is shown in Fig. 4 (a). The far field are the intensity values at the spherical surface originated from the OPA (x=0, y=0) with a radius of 1m and (ϕ, θ) represent the angles from z-axis in x-z and y-z planes respectively. A special case with intensity v.s. θ at ϕ=0 shows one of the directions with the worst side lobe suppression which generally limits this OPA system to about 60° field of view (FOV). A typical far-field, antennae phase map, and far-field intensity with θ at φ=0 with randomly distributed initial phase between 0 and 2π are shown Fig. 4 (d) (e) (f) respectively. The far field intensity is scaled with the same range as in the ideal case. It can be easily seen that a widely distributed initial phase errors generally destroy constructive interference and result in chaotic far field and much lower peak values. The same three plots after the initial phase errors are calibrated by our modified REV method are shown in Fig. 4 (g) (h) (i). The calibration is processed in one pass with 8-bit phase tuning resolution and Gaussian intensity error as mentioned earlier. It can be seen that the calibrated results agree very well with the ideal phase condition results and the system is considered to be calibrated (although the initial phases at some antennae may be off from their ideal values).

The 1D far field in Fig. 4 (c) may be easily use as a visual benchmark to compare calibration results among cases while it is more convenient to use the relative intensity, i.e. scaled with the same range as the ideal case, at Q (θ = 0, φ = 0) as a figure-of-merit to quantitatively compare the statistical calibration performance by different methods at different conditions. We also call it calibrated intensity in following texts.

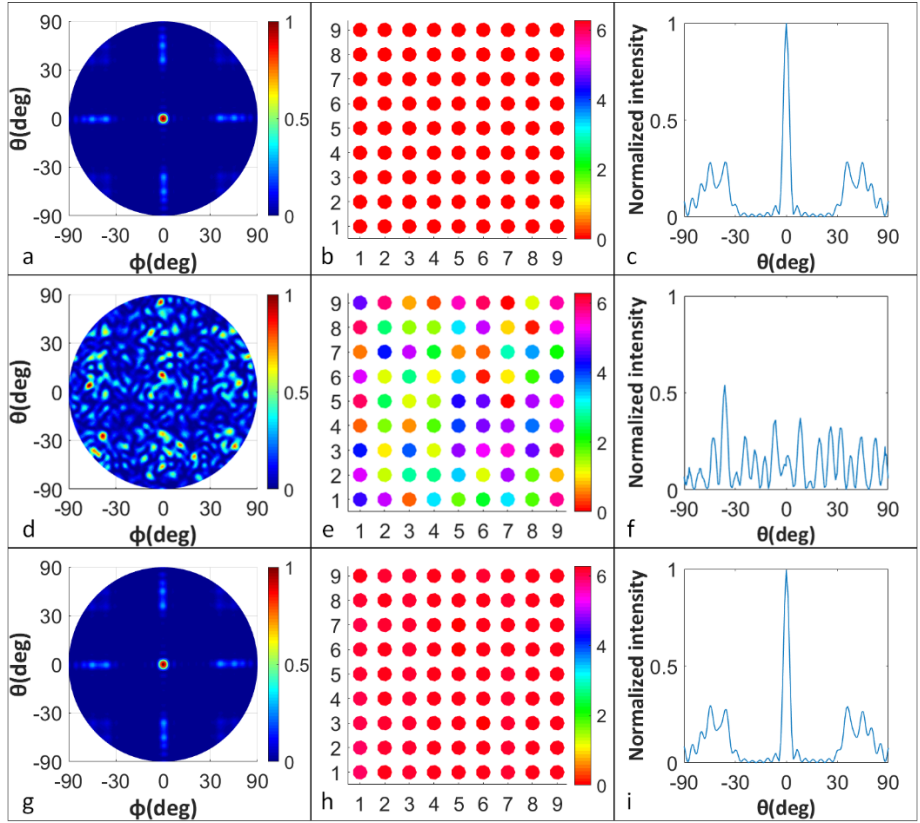

Fig. 4 (a) (b) (c) Far-field, antennae phase map, and far-field intensity with θ at φ=0, respectively, at ideal (equal for all antennae) initial phase condition; (d) (e) (f) Far-field, antennae phase map, and far-field intensity with θ at φ=0, respectively, with randomly distributed phase errors; (g) (h) (i) Far-field, antennae phase map, and far-field intensity with θ at φ=0, respectively, after the initial phase errors are calibrated by our modified REV method.

To compare the calibration performance between our modified REV (mREV) and traditional REV as well as to test the effect on phase tuning resolution, we performed calibration process by both mREV and REV for 1000 cases of randomly generated (within 0-2π) initial phase maps with phase tuning resolutions ranging from 3-bit to 8-bit. The results are summarized in Fig. 5 (a). It can be seen that at any phase tuning resolution the overall calibration performance of mREV is much better than REV in terms of the average calibrated intensity and its error (i.e. root-mean-square error). To show how the calibrated intensity statistically distributes, we pick the 6-bit phase resolution case and plot the probability distribution of calibrated intensity of the 1000 cases processed by mREV and REV, respectively, in Fig. 5 (b). It clearly shows that the calibrated intensity of REV is less accurate and more diverged than that of mREV. In practice, it's obviously preferred to have a robust product calibration process with results produced by mREV.

For mREV, the increase of average calibrated intensity with phase tuning resolution, i.e. number of bits, is easy to understand as the accuracy of $\Delta\phi$ for each antenna increases. The same analysis doesn't apply to the REV cases as shown in Fig. 5 (a) where the average calibrated intensity is slightly higher at lower resolution while the maximally achieved calibrated intensity (upper boundary) presents at medium resolutions. As we already know for REV, another source of error, π phase error in certain initial phase conditions, sometimes determines overall calibration performance. At low resolution, i.e. smaller bits, although the phase accuracy is low the chance to make a π error is also low as the intensity difference between two adjacent phase tuning steps is often too large to produce incorrect determination even with the presence of power measurement errors.

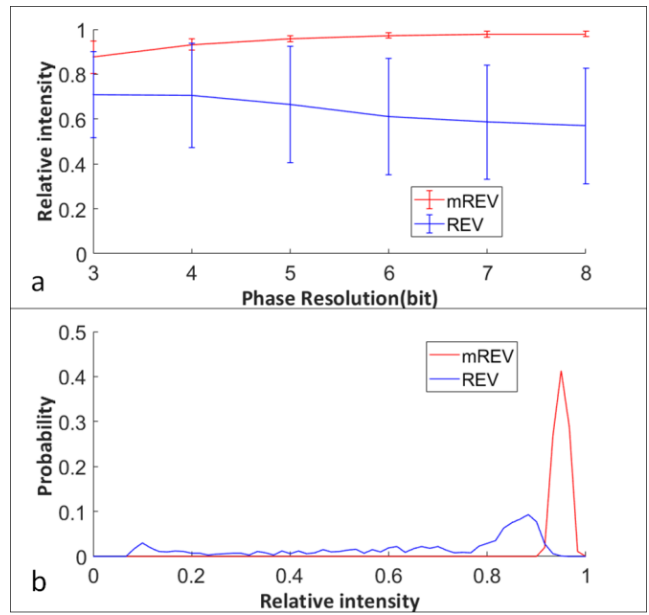

Fig. 5 (a) Statistical results (average and rms error) of the calibrated intensity at Q as a function of phase tuning resolution by our modified REV (mREV) and traditional REV respectively; (b) The probability distribution of calibrated intensity by mREV and REV, respectively, at 5-bit phase resolution.

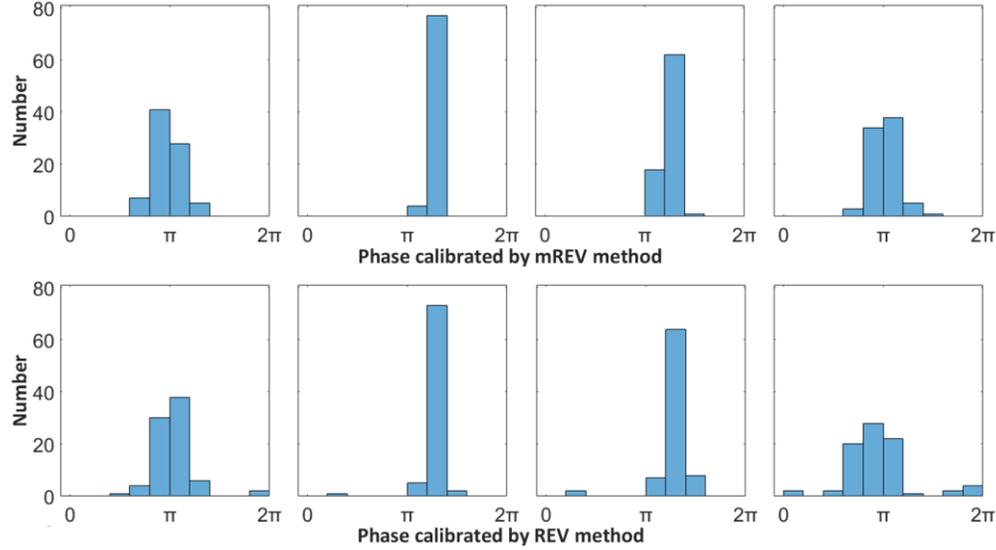

Fig. 6 Histograms of the calibrated antennae phases in typical cases calibrated by mREV and REV methods, respectively, at 6-bit phase resolution.

The above analysis explains the histograms of the calibrated antennae phases when we closely examine the results. Some typical histograms are shown in Fig. 6. The calibrated phases are generally more concentrated within a narrow range in mREV cases. While in cases calibrated by traditional REV, although phases of many antennae are also within a similar narrow range there are always certain phases are located in another π-shifted range and such errors cause poor calibration result.

## 4. Conclusion

The paper presents a modified rotating element electric field vector (modified REV) method for calibrating the antenna array initial condition of an optical phased array (OPA) with arbitrary initial phase distribution, the number of antennae or the arrangement of the antennae. We prove in numerical simulation that such method results in statistically more accurate and predictable calibration compared to traditional REV method which suffers from possible π phase error when in reality wide initial phase distribution and finite optical power measurement accuracy present. We believe the presented method provides great benefits in practice when calibrating large number of OPA products, especially made by integrated photonics, which demands solid and predictable calibration.

## Funding

National Natural Science Foundation of China (Grant No. 61475188, 61705257, 61635013 and 61675231); Strategic Priority Research Program of the Chinese Academy of Sciences (Grant No. XDB24030600).

## Acknowledgments

The authors would like to thank SiQi Li, XingYi Li in the Xi'an Institute of Optics and precision Mechanics (Xi'an, China) for their help.